      [1999/12/01 v1.4c Il Nuovo Cimento]
\documentclass{cimento}

\usepackage{graphicx}  
\title{A Review of Early-Time Optical Follow-ups with 2-m Robotic Telescopes}
\author{A.~Gomboc\from{ins:x}\from{ins:y}\ETC,
C.~Guidorzi\from{ins:y}\thanks{Current address: INAF - Osservatorio Astronomico di Brera, Merate (LC), Italy},
C.~G.~Mundell\from{ins:y}, A.~Melandri\from{ins:y},\\
A.~Monfardini\from{ins:y}, D.~Bersier\from{ins:y}, M.~F.~Bode\from{ins:y},
D.~Carter\from{ins:y}, S.~Kobayashi\from{ins:y}, C.~J.~Mottram\from{ins:y}, R.~J.~Smith\from{ins:y} 
\atque I.~A.~Steele\from{ins:y} on behalf of the {\it RoboNet-1.0} collaboration
}

\instlist{\inst{ins:x} Faculty of Mathematics and Physics, University of Ljubljana, Jadranska 19, 1000 Ljubljana, Slovenia
  \inst{ins:y} Astrophysics Research Institute, Liverpool John Moores University, Twelve Quays House, Egerton Wharf, Birkenhead, CH41 1LD, UK}
\PACSes{\PACSit{01.30.Cc}{Conference proceedings}
\PACSit{95.55.Cs}{Ground-based ultraviolet, optical and infrared telescopes}
\PACSit{98.70.Rz}{gamma-ray sources; gamma-ray bursts}}

\begin{document}

\maketitle

\begin{abstract}
We summarise recent deep, rapid GRB follow-up observations using the {\it RoboNet-1.0} network which comprises three fully-robotic 2-m telescopes, 
the Liverpool Telescope and the Faulkes Telescopes North and South.
Observations begin automatically within minutes of receipt of a GRB alert and may continue for hours or days to provide well-sampled multi-colour 
light curves or deep upper limits. Our light curves show a variety of early afterglow behaviour, from smooth, simple or broken power laws to 'bumpy', 
for a wide range of optical brightness (from the unprecedented faint detections of GRB~060108 and GRB~060510B to classical bright ones).
We discuss GRB~051111 as an example of how the combination of optical and X-ray light curves can provide insight into the circumburst environment, 
in particular the role played by intrinsic extinction soon after the burst.

\end{abstract}

\section{{\it RoboNet-1.0}}
{\it RoboNet-1.0} is a network of three 2-m robotic telescopes, situated around the globe: the Liverpool Telescope (LT) on La Palma, Canary Islands, 
Faulkes Telescope North (FTN) on Mauna Kea, Hawaii, and Faulkes Telescope South (FTS) in Siding Spring, Australia. 
They are three identical telescopes (2-m aperture, fully opening enclosure) 
which were built by Telescope Technologies, Ltd, adjacent to the LJMU Astrophysics Research Institute, which is operating the LT. 
The FTs were initially funded by the Dill Faulkes Educational Trust, and are currently funded by Las Cumbres Observatory with most of their observing time 
intended for educational purposes. Access to research time on all three large robotic telescopes is the core of the {\it RoboNet-1.0} project, 
which is led by LJMU, funded by UK PPARC, and includes members of 9 other UK university teams in Cardiff, Exeter, Hertfordshire, Leicester,
Manchester, Mullard Space Science Laboratory, Queen's University Belfast, St.~Andrews and Southampton. 
The integration of the LT and FTs gives 
an increased sky and time coverage in rapid response to GRB alerts and optimised monitoring of GRB afterglows. 

\subsection{Instrumentation}
All three telescopes have five instrument ports with permanently mounted instruments, which can be selected within 30 sec. They all have optical 
CCD camera with 4.6'$\times $4.6' field of view and several filters (B, V, g', r', i', z' on LT; U, B, R, i', H$\alpha$, OIII on FTs),
and can typically reach limiting magnitudes r', R=19 in 10 sec exposure.
They also have low resolution spectrographs: FTs' were developed by the University of Leicester in collaboration with University of Manchester and LJMU,
have 4400-8500~\AA \, range at a resolution of 4~\AA \, and give S/N=10 for V=15 in 100 sec exposure. For details on LT spectrographs see~\cite{ref:gom}. 
In addition, the LT's instrumentation includes infrared camera (see~\cite{ref:gom}) and recently developed RINGO polarimeter. 
It is the ring polarimeter based on the 
design by Clarke \& Neumayer~\cite{ref:ringpol}. It is designed to give polarimetric accuracy $<$ 1\% for 15 mag object in 10 min exposure time. For details see~\cite{ref:steele}.

\subsection{GRB observing strategy and LT TRAP}
Fully robotic operation of all three {\it RoboNet-1.0} telescopes enables rapid response to GRB alerts, typically starting observations in 2-5 min after the trigger. Together 
with 2-m aperture this results in  early {\it and} deep observations, which are crucial in detecting optically faint afterglows and in setting deep upper limits in case of optically dark GRBs.

In case of an optical afterglow detected, the range of filters allows the study of multi-colour light curves evolution and the variety of early afterglows. 
Combined with
data from Swift BAT, XRT and UVOT it allows the construction of early Spectral Energy Distribution, which is vital in the study of GRB 
physics 
and circumburst environment.

In case of exceptionally bright afterglow (R$<$15)  early spectrometry and polarimetry (on LT) may help discern among different models of the GRB explosion.

GRB observing strategy on {\it RoboNet-1.0} telescopes is fine tuned to each telescope's instrumentation and is described in more detail in~\cite{ref:guiPASP}.
Crucial part of the automatic follow-up observations is the Liverpool Telescope Transient Rapid Analysis Pipeline (LT TRAP, also described in~\cite{ref:guiPASP}), 
which in real time identifies OT candidates, assigns confidence level to them,
estimates their magnitudes and creates and up-dates in real time a light curve for each candidate. 

\section{GRBs observed to date}
Since the implementation of fully automated GRB follow-up on {\it RoboNet-1.0} telescopes in early 2005, until now (Oct 2006) we have responded to more than 30 GRB alerts. 

\subsection{Early and deep upper limits}
In roughly half of the cases we did not detect any optical afterglow, but we could put early upper limits on the possible afterglow: typically our limits are 
R, r' $\sim$ 19  at $\sim$ 5 min and $\sim$21 at $\sim$20 min after the trigger time.

\subsection{Optical afterglows detected}
\begin{figure}
\begin{center}
\includegraphics[height=0.58\textwidth]{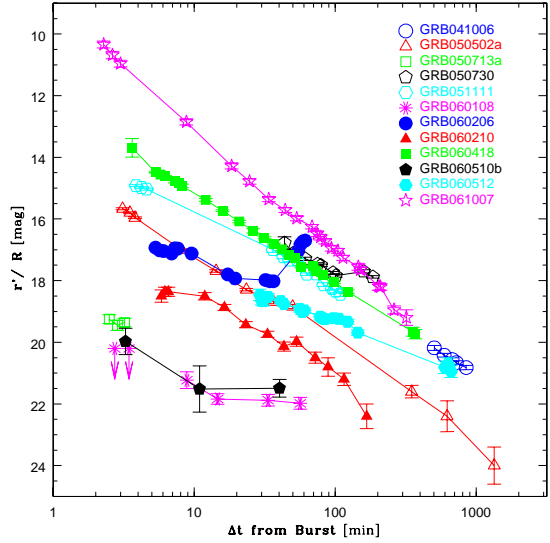}  
\caption{A sample of GRB afterglow light curves observed by the LT in r' band (GRB~041006, GRB~050502A, GRB~050713A, GRB~050730, GRB~060206, GRB~060418, GRB~060512) 
and by the FTN and FTS in R band (GRB~051111, GRB~060108, GRB~060210, GRB~060510B, GRB~061007).}
\label{gomfig1}
\end{center}
\end{figure}
The first early ($<$1 hr) multi-colour light curve of an optical afterglow was obtained by the LT in case of GRB~050502A~\cite{ref:guiApJ}.
Afterglows that we detected and followed up since then, show a variety of light curve behaviour (see Fig.~\ref{gomfig1}): from monotonous power-law 
decay (e.g. GRB~060418, GRB~061007, for details see~\cite{ref:mun}), 
broken power-laws (GRB~051111), small bumps (GRB~060512) 
to large rebrightening as in the case of GRB~060206 (for details see~\cite{ref:gui,ref:mon}).

\begin{figure}
\begin{center}
\includegraphics[height=0.44\textwidth]{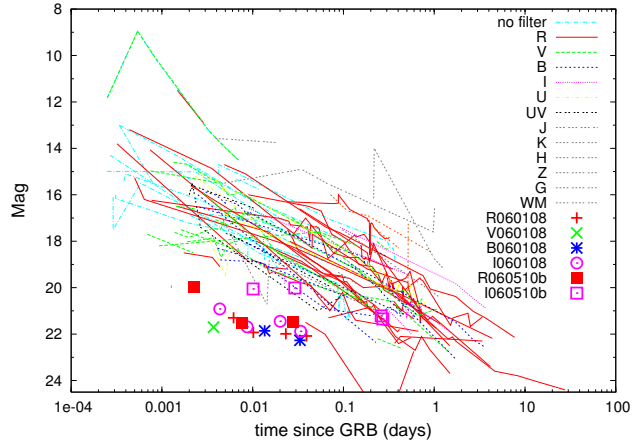}  
\caption{Early optical afterglows and two very faint ones, detected by {\it RoboNet-1.0}: GRB~060108 and GRB~060510B.}
\label{gomfig2}
\end{center}
\end{figure}

We also detected two cases of very faint optical afterglows, GRB~060108 and GRB~060510B (Fig.~\ref{gomfig2}). As discussed in ~\cite{ref:gui,ref:oat}, 
the faintness of the former is due to combination of intrinsic optical faintness and significant suppression by dust, while the faintness of the 
latter is due to high redshift of z=4.9~\cite{ref:pri}.

\section{Circumburst environment of GRB~051111}
The Swift detection of a GRB~051111 (prompt $\gamma$-ray emission showed a classical FRED profile) triggered early BVRi' observations with the FTN 
starting 3.9 min after the trigger time. Results are described in more detail in~\cite{ref:guiAA}, here we give only a brief summary: 
The optical light curves are well fitted with a broken power-law ($\alpha_1$=0.35, $\alpha_2$=1.35, $t_{break}$=12 min).
Simultaneous Swift XRT observations allowed the construction of IR to X-ray Spectral Energy Distribution at 
80 min after the burst, which is consistent with the cooling break lying between the optical and X-ray bands. Our modelling of the circumburst environment 
suggests dust with big grains or grey extinction profiles. We also find evidence for an overabundance
of some $\alpha$ elements such as oxygen or, alternatively, for a significant presence of molecular gas.

\section{Conclusion}

The rapid and accurate localisation of GRBs by the Swift satellite has revolutionised ground-based follow-up observations  
and enable uncovering of previously unknown behaviour (e.g. long-lived central engines and unexplained chromatic breaks), 
that is confounding established GRB fireball models. 
As also observations by {\it RoboNet-1.0} telescopes reveal, early and later-time optical light curves are also more complex than previously thought,  
challenging the basic scenario of a smooth power law decay followed by a 
jet break.  
To fully understand GRB physics, their environment and progenitors, well-sampled light curves across a wide wavelength band and 
time range are the key.


\acknowledgments

We acknowledge valuable collaboration with colleagues at the Universities of Leicester, Hertfordshire and Mullard Space Science Laboratory.
The Liverpool Telescope is operated by 
Liverpool John Moores University with support from the UK PPARC. The Faulkes Telescopes were operated with support from the Dill 
Faulkes Educational Trust and the Las Cumbres Observatory Global Telescope.

\end{document}